\documentclass{article}
\usepackage{ijcai22}

\usepackage{times}

\usepackage{soul}
\usepackage{url}
\usepackage[hidelinks]{hyperref}
\usepackage[utf8]{inputenc}
\usepackage[small]{caption}
\usepackage{graphicx}
\usepackage{amsmath}
\usepackage{booktabs}
\usepackage{float}
\usepackage{multirow}
\usepackage{diagbox}
\usepackage[linesnumbered,titlenumbered,ruled,vlined,commentsnumbered,noend]{algorithm2e}

\usepackage{xcolor}
\usepackage{comment}
\usepackage{color}

\usepackage{epsfig}
\usepackage{graphicx}
\usepackage{amsmath}
\usepackage{amssymb}

\usepackage{multirow}
\usepackage{enumitem}
\usepackage{moresize}
\usepackage{epstopdf}
\usepackage{array}

\usepackage{makecell}
\usepackage{stfloats}

\usepackage{wrapfig}
\usepackage{subfigure}
\usepackage[hang,flushmargin]{footmisc} 

\def\minimize{\operatornamewithlimits{minimize}}

\usepackage{xspace}

\let\OLDthebibliography\thebibliography
\renewcommand\thebibliography[1]{
  \OLDthebibliography{#1}
  \setlength{\parskip}{1.6pt}
  \setlength{\itemsep}{0pt plus 0.3ex}
}

\newcommand{\etal}{\textit{et al}.{ }}

\newcommand{\eg}{\textit{e}.\textit{g}., }
\newcommand{\etc}{\textit{etc}.{ }}

\definecolor{darkorange}{rgb}{1.0, 0.55, 0.0}
\definecolor{lincolngreen}{rgb}{0.11, 0.35, 0.02}
\definecolor{cornflowerblue}{rgb}{0.39, 0.58, 0.93}
\definecolor{cobalt}{rgb}{0.0, 0.28, 0.67}

\title{\emph{Anti-Forgery}: Towards a Stealthy and Robust DeepFake Disruption Attack via Adversarial Perceptual-aware Perturbations}

\author{
Run Wang$^{1,2}$\footnote{Run Wang’s Email: \href{mailto:wangrun@whu.edu.cn}{wangrun@whu.edu.cn}}\and
Ziheng Huang$^{1,2}$\and 
Zhikai Chen$^3$\and
Li Liu$^4$\and
Jing Chen$^{1,2}$\and
Lina Wang$^{1,2}$
\affiliations
$^1$School of Cyber Science and Engineering, Wuhan University, China  $^2$Key Laboratory of Aerospace Information Security and Trusted Computing, Ministry of Education, China \\ $^3$Tencent Zhuque Lab  $^4$Fudan Development Institute, Fudan University, China\\
}

\begin{document}

\maketitle

\begin{abstract}

DeepFake is becoming a real risk to society and brings potential threats to both individual privacy and political security due to the DeepFaked multimedia are realistic and convincing. However, the popular DeepFake passive detection is an ex-post forensics countermeasure and failed in blocking the disinformation spreading in advance. To address this limitation, researchers study the proactive defense techniques by adding adversarial noises into the source data to disrupt the DeepFake manipulation. However, the existing studies on proactive DeepFake defense via injecting adversarial noises are not robust, which could be easily bypassed by employing simple image reconstruction revealed in a recent study MagDR~\cite{chen2021magdr}. 

In this paper, we investigate the vulnerability of the existing forgery techniques and propose a novel \emph{anti-forgery} technique that helps users protect the shared facial images from attackers who are capable of applying the popular forgery techniques. Our proposed method generates perceptual-aware perturbations in an incessant manner which is vastly different from the prior studies by adding adversarial noises that is sparse. Experimental results reveal that our perceptual-aware perturbations are robust to diverse image transformations, especially the competitive evasion technique, MagDR via image reconstruction. Our findings potentially open up a new research direction towards thorough understanding and investigation of perceptual-aware adversarial attack for protecting facial images against DeepFakes in a proactive and robust manner. We open-source our tool to foster future research. Code is available at \href{https://github.com/AbstractTeen/AntiForgery/}{\textcolor{blue}{https://github.com/AbstractTeen/AntiForgery.}}
\end{abstract}
\section{Introduction}\label{sec:intro}
In recent years, we have witnessed the remarkable development of GAN in image synthesis and fine-grained image manipulation. Attackers could leverage GAN to generate realistic and natural synthetic images, audios, and videos (also known as DeepFake), which poses potential security and privacy concerns to individuals~\cite{juefei2021countering,wang2020deepsonar}. In this AI era, we are living in a world where we cannot believe our eyes and ears anymore. Thus, effective countermeasures should be developed for fighting against DeepFakes.

To defend against DeepFakes, various countermeasures are developed in both passive and proactive manner. However, both of them are still in their early stage and not prepared for tackling this emerging severe threat. The passive DeepFake detection merely determines the real or fake which is an ex-post DeepFake defense manner~\cite{hu2021exposing}. More importantly, the existing DeepFake detection techniques can not well tackle the DeepFakes created by unknown synthetic techniques~\cite{wang2021faketagger}. According to a result of the DeepFake detection challenge (DFDC) held by Facebook, the winner team can give a detection accuracy less than 70\%. Thus, some researchers are working on developing proactive DeepFake defending techniques by adding adversarial noises into the source images to disrupt the DeepFake creation~\cite{yang2020defending,segalis2020disrupting}. Specifically, they hope that the created DeepFakes with added perturbations exhibit visually noticeable artifacts and provide signals for detectors. The DeepFake disruption is the most promising countermeasures for fighting against DeepFakes in a proactive manner, which shows potential for tackling DeepFakes in the wild~\cite{chen2021magdr}. However, existing DeepFake disruption studies by adding adversarial noises suffer challenge in tackling input transformations, which limits their practicality applications.

Existing studies mostly borrow the idea of prior adversarial attack (\eg{}FGSM, PGD) via gradient-based or optimization-based strategies to generate imperceptible adversarial noises. However, a series of studies reveal that such adversarial noises could be easily removed or destroyed. A recent study, MagDR~\cite{chen2021magdr}, illustrated that a simple input reconstruction could destroy the added adversarial noises for disrupting DeepFakes. 

The ultimate goal of proactive DeepFake defense is to create low-quality DeepFakes when applying various techniques for forgery purpose. The created low-quality DeepFakes will exhibit noticeable artifacts and could be easily spotted even with simple DeepFake detectors. Thus, a desired proactive DeepFake defense technique should satisfy the following three properties, \textit{robust} to input transformations (\eg{} reconstruction), visually \textit{natural} perturbations to human eyes, working on the \textit{black-box} settings without obtaining any knowledge of the forgery model.

\begin{figure*}[t]
\centering
\includegraphics[width=1\linewidth]{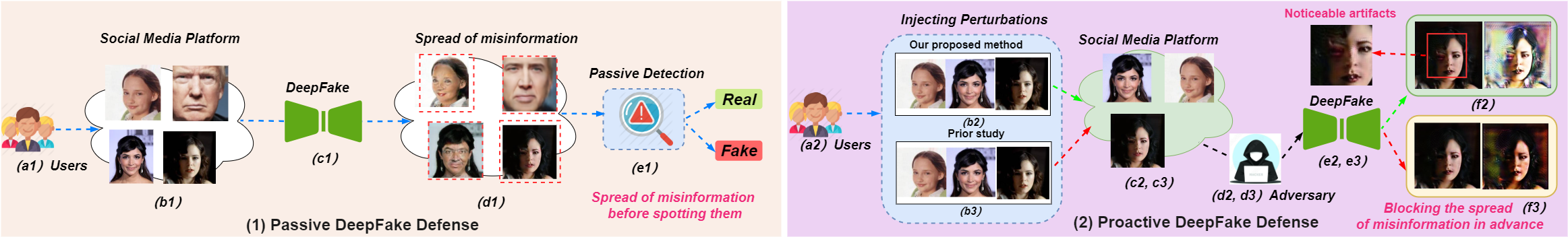}
\caption{An overview of the difference between passive and proactive DeepFake defense. The passive DeepFake detection failed in blocking the spread of misinformation before the fake is confirmed, while the proactive DeepFake defense provides a restricted spread of misinformation as the noticeable artifacts exhibited in the created DeepFakes. In contrast to the prior study working on RGB color space, our proposed anti-forgery method operating on the Lab color space could survive the input transformation well with noticeable artifacts to block the further spread of misinformation.}
\label{fig:overview}
\end{figure*}

In this paper, we investigate the vulnerability of GANs in image synthesis and seek a kind of robust adversarial perturbation which could survive drastic input transformation. Specifically, we propose a novel anti-forgery approach by adding perceptual-aware perturbations into the source data to enforce the created DeepFakes exhibiting noticeable artifacts and detectable with simple DeepFake detectors. Our perceptual-aware perturbations operate on the \textit{Lab} color space in the de-correlated $a$ and $b$ channel in an incessant manner. Unlike the prior studies working on the \textit{RGB} color space which introduces perceivable distortions incurred by unnatural colors even with small variations~\cite{huang2021initiative}, our proposed method works on a natural-color range of semantic categories which generates perceptually natural facial images. More importantly, our generated perceptual-aware perturbation is robust against input transformation which shows potential to be deployed in the real world. Figure~\ref{fig:overview} illustrates the comparison with the existing common DeepFake disruption techniques by adding adversarial noises on the RGB color space.

Extensive experiments conducted on the three types of DeepFake demonstrate the effectiveness of our proposed method and robustness against the input transformation, like reconstruction, compression. Additionally, some ablation studies are conducted for illustrating that our added perturbations could provide a clear signal for DeepFake detection. Our main contributions are summarized as follows:

\begin{itemize}[leftmargin=*]
\item We introduce a novel anti-forgery method by adding adversarial perceptual-aware perturbations against forgery in a proactive manner by transforming facial images in an incessant and natural way, as opposed to adding meaningless and sparse noises in prior studies which are not robust to drastic input transformations, like reconstruction.
\item We employ a simple yet effective method to generate the perceptual-aware perturbations by operating on the \emph{Lab} color space in the decorrelated $a$ and $b$ channel that generates a visually natural face which could be further leveraged disrupting the DeepFake creation.
\item For the first time, we conduct experiments on three types of DeepFakes to demonstrate the effectiveness of our devised perceptual-aware perturbations in creating DeepFakes with noticeable artifacts and robustness in surviving the input reconstruction compared with prior studies.
\item Our research findings hint a new research direction towards perceptual-aware adversarial attack by investigating natural and robust perturbations against various transformations. We hope that more studies working on exploiting the vulnerabilities of GAN in image synthesis to develop DeepFake proactive countermeasures which could fight against DeepFakes in the wild.
\end{itemize}

\section{Related Work}\label{sec:related}
\subsection{DeepFake Creation}
In the past years, GAN and its variants have achieved tremendous success in image synthesis and manipulation. DeepFake leverages the success of GANs to generate forgery facial images or videos, which pose potential threats to individual privacy and political security. We are living in a world where we cannot believe our eyes anymore. In general, \textit{entire synthesis}, \textit{attribute editing}, \textit{identity swap}, and \textit{face reenactment} are four common types of DeepFakes.

The entire synthesis produces non-existent images in the world with random vectors to the network, such as PGGAN, StyleGAN. The attribute editing is a fine-grained face manipulation by modifying the simple face attributes (\eg hair color, bald) and complex face attributes (\eg gender, age) with popular GANs, like STGAN~\cite{liu2019stgan}, StarGAN~\cite{choi2018stargan}. The identity swap is known as the popular face swap by swapping the face between target and source facial image. FaceSwap\footnote{https://github.com/Oldpan/Faceswap-Deepfake-Pytorch} and DeepFaceLab~\cite{perov2020deepfacelab} are two popular available tools for identity swap. Similar to identity swap, face reenactment also known as expression swap involves facial expression swap between target and source facial image by applying popular tools, such as Face2Face~\cite{thies2016face2face}.

Among the four types of DeepFake, the entire synthesis produces non-existent facial images without involving any source image manipulation, thus it will not bring any privacy or security concerns to individuals. In this paper, we mainly focus on the other three types DeepFakes (\eg attribute editing, identity swap, and face reenactment), which pose security and privacy concerns to individuals. 

\subsection{DeepFake Defense} 
\textbf{DeepFake Detection}. Determining if the facial image is real or fake by observing the subtle differences between real and fake images is a straightforward idea for defending DeepFakes. However, DeepFake detection is an ex-post forensic manner and suffers various known challenges, such as generalization to tackle unknown synthetic techniques, robust against quality degradations, evasion to adversarial attacks, which are obstacles for its practical usage to be deployed in the wild~\cite{juefei2021countering}.

Studies have shown that the subtle differences between real and fake could be revealed in the spatial~\cite{wang2020cnn,wang2021fakespotter} and frequency domain~\cite{qian2020thinking}, however, they are susceptible to adversarial attack via various image-level and frequency-level manipulations~\cite{carlini2020evading,huang2020fakepolisher}. Another line work investigate the biological signals which are hard to replicate~\cite{hu2021exposing}. Unfortunately, these passive DeepFake detection techniques cannot be applied for blocking the wide spreading of DeepFakes in advance before causing damage impacts.

\noindent\textbf{DeepFake Disruption}. To address the aforementioned limitations of DeepFake detection, DeepFake disruption aims at disrupting the images being DeepFaked proactively by adding adversarial perturbations into the source image to produce damaged images. Recently, massive studies have shown that Deep Neural Networks (DNN) are vulnerable to adversarial attack (\eg FGSM, PGD) by adding imperceptible noises to craft so-called adversarial examples. Thus, some researchers explore whether the DeepFakes are vulnerable to adversarial examples as well. 

Yang \etal{}leveraged differentiable random image transformations to yield adversarial faces with noticeable artifacts~\cite{yang2020defending}. Ruiz \etal{}presented a spread-spectrum adversarial attacks against conditional image translation network in a grey-box scenario~\cite{ruiz2020disrupting}. Yeh \etal{}proposed two types of adversarial attack against image translation GANs with designed adversarial loss function by gradient optimization to output blurred and distorted output~\cite{yeh2020disrupting}. Huang \etal{}\cite{huang2021initiative} proposed an initiative defense framework against facial manipulation by injecting venom into the target manipulation model in the black-box settings. This work adds invisible perturbations by training a surrogate model, however, they merely claim its effectiveness in attribute editing and face reenactment since these two types of manipulation model share a similar pipeline. Unfortunately, it will be hardly applied in defending identity swap which is more challenge to find an appropriate surrogate model to generate perturbations.

Unfortunately, the existing DeepFake disruption techniques employ a naive adversarial attack to generate invisible perturbations which could be easily detected and corrupted by input reconstruction. A recent study MagDR~\cite{chen2021magdr} also demonstrates this and reveals that the DeepFake disruption techniques by adopting C\&W and PGD are all failed in producing damaged images. Specifically, MagDR employs image restoration techniques to remove the added perturbations and obtains the desired form further.

In this paper, we challenge the effectiveness of MagDR in dealing with our perceptual-aware perturbations which are generated in an incessant and natural manner. We aim at developing stealthy and robust imperceptible adversarial perturbations which could effectively survive MagDR and provide strong protection to our shared faces in social media.

\section{Problem Statement}\label{sec:problem}
In this paper, we describe the DeepFake defense from both passive DeepFake detection and proactive DeepFake defense in Figure~\ref{fig:overview}. In the real world, a user uploads his/ her personal facial images to social media, like Twitter to share with friends or anyone on the Internet. Unfortunately, an attacker could easily pick the victim's photos for malicious manipulation with GAN to create fake pornography for individuals to raise privacy concerns or to generate a fake official statement for celebrities to cause panic. Our method adds adversarial perceptual-aware perturbations into personal images before uploading to social media without introducing any visual artifacts. The facial image with our applied perturbations could prevent the GAN-based manipulations well by generating damaged facial images with noticeable artifacts. The key idea here is that our perturbed facial images should be robust enough to survive input transformations, like reconstruction before the DeepFake process. Finally, the proactive DeepFake defense is achieved for creating damaged images and exhibiting clear fake signals for detectors.

Here, we elaborate the details regarding Figure~\ref{fig:overview}. In the left panel, we present the pipeline of passive DeepFake detection. A user (Figure~\ref{fig:overview}-a1) first uploads personal facial images to the social media platform (Figure~\ref{fig:overview}-b1). Then, attackers pick the shared photos to create DeepFake (Figure~\ref{fig:overview}-c1) (\eg{}change the color of hair, the intensity of facial expressions) which causes the spread of misinformation in the platform (Figure~\ref{fig:overview}-d1). Finally, the DeepFake detector (Figure~\ref{fig:overview}-e1) determines whether the suspicious one is real or fake in an ex-post forensics manner to further block the misinformation spreading in social media platforms if fake is confirmed. In the right panel, we present the pipeline of proactive DeepFake defense by comparing our proposed method with the prior DeepFake disruption approach by employing popular adversarial noises.

In contrast to the passive DeepFake detection in the left panel, a user first (Figure~\ref{fig:overview}-a2) injects perturbations into facial images before uploading them to social media platforms (Figure~\ref{fig:overview}-c2, c3). Our proposed method injects perceptual-aware perturbations by operating the Lab color space in an incessant manner (Figure~\ref{fig:overview}-b2), in comparing with prior studies inject imperceptible adversarial noises by perturbing the RGB color space in a sparse manner (Figure~\ref{fig:overview}-b3). However, in a real battleground scenario, an adversary (Figure~\ref{fig:overview}-d2, d3) tries to remove or corrupt the added perturbations by employing input transformations (\eg{}reconstruction) before creating DeepFakes. In creating DeepFakes (Figure~\ref{fig:overview}-e2, e3), our proposed method adding perceptual-aware perturbations could survive the adversarial attack with input transformations and exhibit noticeable artifacts in the manipulated facial images (Figure~\ref{fig:overview}-f2) and blocked immediately without further spreading to cause any panic and privacy concerns for individuals. However, the prior studies are not robustly output realistic and natural DeepFakes which cause the spreading of misinformation (Figure~\ref{fig:overview}-f3).

In summary, in such a real battleground scenario, both the passive DeepFake detection and prior proactive DeepFake disruption methods failed in blocking the misinformation spreading before resulting severe social impacts.

\section{Methodology}\label{sec:method}
\subsection{Insight}\label{sec:insight}
DeepFake disruption is a promising countermeasure for defending DeepFakes proactively~\cite{chen2021magdr}. However, the existing studies suffer two issues, 1) they are not robust to the input transformation which is demonstrated in MagDR, 2) the effectiveness on all types of DeepFakes is unclear~\cite{huang2021initiative}, which is an obstacle for its practical usage. Thus, the community is not prepared for tackling this emerging threat.

A straightforward idea for defending GAN-based DeepFake proactively is to explore the vulnerability of GAN in image synthesis and manipulation. To this end, adversarial examples are employed for attacking GAN by adding imperceptible perturbations into the input to introduce distortions on the output. Generally speaking, the created output with perturbations should satisfy the following three properties when defending DeepFakes in the wild. 

\begin{itemize}[leftmargin=*]
    \item Visually natural to human eyes. The uploaded facial images to the social media platform are mainly used to share exciting moments to friends at the most of the time, thus the visually quality degradation is not allowed.
    \item Robust to input transformation (\eg{}reconstruction). In the real world, adversaries will corrupt the added perturbations intentionally to break the defense mechanism. 
    \item Generalizing well to a wide types of DeepFakes. A facial image spreading in the social media platform would suffer various forgery which all cause security and privacy concerns to individuals.
\end{itemize}

Inspired by the advantages of Lab color space that are perceptual uniform and more satisfy the humans' to the outside world, we operate the $a$ and $b$ channel of Lab color space to generate natural color perturbations. More importantly, our perturbations are generated in an incessant manner, unlike the perturbations generated on RGB color space which are sparse and introduce distortions even small variations existed. We illustrate the power of Lab color space in color representation in Section~\ref{sec:diff}. In the following subsections, we introduce the framework of our proposed anti-forgery method to generate adversarial perceptual-aware perturbations to disrupt the DeepFake creation.

\subsection{Adversarial Attack to GAN}
The foundation of DeepFake creation is GAN which consists of two deep learning networks, the Generator $\mathnormal{G}$ and Discriminator $\mathnormal{D}$. In the GAN training, the Generator $\mathnormal{G}$ tries to generate samples indistinguishable to the real, while the Discriminator $\mathnormal{D}$ learns to differentiate synthesized sample $\mathnormal{G(z)}$ from noises $z$ and real samples. For example, a typical image-to-image translation network CycleGAN could be employed for identity swap\footnote{https://github.com/shaoanlu/faceswap-GAN}. Specifically, CycleGAN learns to build a mapping $G: x\to y$ and an inverse mapping $F: y \to x$ between two image domain $x$ and $y$.

Similar to adversarial examples, the imperceptible perturbations used to disrupt DeepFakes are produced by introducing minimal distortion to preserve the natural effects of human eye vision. Let $x$ be the input source image, $\theta$ is the imperceptible perturbation, $\Tilde{x}$ is the crafted input with adversarial perturbations.
\begin{equation} \label{eq:1}
\setlength\abovedisplayskip{3pt}
    \setlength\belowdisplayskip{3pt}
\Tilde{x} = x + \theta
\end{equation}

A generator $G$ receives the two input $x$ and $\Tilde{x}$ and outputs $G(x)$ and $G(\Tilde{x})$ further. Ideally, $G(x)$ and $G(\Tilde{x})$ are totally different where $G(\Tilde{x})$ exhibits noticeable artifacts to human eyes. More specifically, the objective function can be formulated as follows by maximize the distortion between $G(x)$ and $G(\Tilde{x})$.
\begin{equation} \label{eq:2}
\setlength\abovedisplayskip{3pt}
    \setlength\belowdisplayskip{3pt}
\max  \limits_{\theta} \mathcal{L}(G(x+\theta), r), \text{   subject to   } \lVert \theta \rVert_\infty \le \epsilon
\end{equation}
where $\epsilon$ is the maximum magnitude of the perturbation, $\mathcal{L}$ is a distance function for measuring the similarity, $r$ is a ground-truth, $r=G(x)$.

\subsection{Lab Color Space}\label{sec:diff}
The Lab color space is designed to conform with human's sense of color which is perceptual uniform. A light channel $L$ and two color channels $a$ and $b$ consist the Lab color space, where the $L$ channel ranges from black ($0$) to white ($100$) representing the light, $a$ channel ranges from green ($-128$) to red ($+127$), $b$ channel ranges from blue ($-128$) to yellow ($+127$). Unlike RGB color space, we can change the lightness by simply modifying the value of $L$ channel without involving any changes to the color channel $a$ and $b$.

Due to the power of Lab color space in representing color space that is perceptual uniform and the wide value range in representing color, in this paper, our adversarial perturbations are generated by operating the Lab color space to add perturbation into the channel $a$ and $b$.

\subsection{Proposed Anti-forgery Method}\label{sec:framework}
Specifically, Algorithm~\ref{alg:procedure} describes the details of our proposed method. Our goal for the generated perturbations is natural to human eyes without introducing any quality degradation and resistant enough to potential input transformation attacks. First, we convert the input image from the RGB color space to the Lab color space for adding perceptual uniform perturbations into the channel $a$ and $b$. Then, the perturbations is updated by attacking the surrogate model $M$ via optimization-based strategy adopted in C\&W.

To improve the transferability in tackling multiple facial attributes, we select different facial attribute label $c$ in each iteration. The objective function can be formulated as follows. 
\begin{equation} \label{eq:2}
\setlength\abovedisplayskip{3pt}
    \setlength\belowdisplayskip{3pt}
\minimize{}\mathcal{L}(M(x_{adv}, c), o)
\end{equation}
where $\mathcal{L}$ could be $L_1$ or $L_2$ norm, $o$ could be $0$, $1$, or even Gaussian noises. When $o$ is a regular translation image, the objective can be formulated as follows.
\begin{equation} \label{eq:3}
\setlength\abovedisplayskip{3pt}
    \setlength\belowdisplayskip{3pt}
\minimize{}-\mathcal{L}(M(x_{adv}, c), o)
\end{equation}

\begin{algorithm}[t]
	\footnotesize
	\SetAlgoLined
	\SetKwInOut{Input}{Input}
	\SetKwInOut{Output}{Output}
	\Input{Input image $x$, Surrogate model $M$, Label $c \in C$, Iteration $K$, Objective $o$, Learning rate $\tau$.}
	\Output{Adversarial sample $x_{adv}$ with perturbation.}
	initialization $\theta_a$, $\theta_b$\\
	\For{$i \in \{1...K\}$}{
	$l, a, b \leftarrow rgb2lab(x)$\\
    \textbf{$\triangleright$ Add perturbations for both channel $a$ and $b$.}\\
	$a^\prime \leftarrow a + \theta_a$\\
	$b^\prime \leftarrow b + \theta_b$\\
    \textbf{$\triangleright$ Convert into RGB color space.}\\
	$x_{adv} \leftarrow lab2rgb(l, a^\prime, b^\prime)$\\
    \textbf{$\triangleright$ Update the perturbations of channel $a$ and $b$.}\\
	$\theta_a \leftarrow \theta_a - \tau\cdot\nabla_{\theta_a} \mathcal{L}(M(x_{adv}, c), o)$\\
	$\theta_a \leftarrow clip(\theta_a, -\epsilon, \epsilon)$\\
	$\theta_b \leftarrow \theta_b - \tau\cdot\nabla_{\theta_b} \mathcal{L}(M(x_{adv}, c), o)$\\
	$\theta_b \leftarrow clip(\theta_b, -\epsilon, \epsilon)$
	}
	\Return $x_{adv}$
	\caption{Our proposed adversarial perceptual-aware perturbations.}
	\label{alg:procedure}
\end{algorithm}

\subsection{Comparison with Popular Adversarial Noises}
A straightforward idea for attacking GAN to disrupt the DeepFake creation would be employing the gradient-based and optimization-based strategies to generate imperceptible perturbations. However, such restricted perturbations operating on the RGB color space suffer sparse issues which could be easily detected and corrupted by input transformations~\cite{meng2017magnet}. Specifically, the three channels in RGB color space have strong correlations, indicating that natural perturbations require the modification of these three channels simultaneously. Thus, the perturbations generated by operating the RGB color space are not robust against real world input transformations.

In summary, our perceptual-aware adversarial perturbations by operating the Lab color space have the following main strengths. 1) Robust to input reconstruction. Our perturbations operated on Lab color space which is perceptual uniform, as opposite to the prior studies operating on the RGB space which is sparse. The perceptual uniform perturbations operated on the Lab color space is more likely to share the same distribution as the unperturbed input. 2) Generic to a wide types of fogery. There is a wide range of color values for operating on the Lab color space. We will always explore the best perturbations for disruption effectively.

\section{Experiments}\label{sec:exp}

\subsection{Experimental Setup} \label{sec:setup}
\textbf{Dataset.} All our experiments are conducted on a popular face dataset CelebFaces Attributes (CelebA)~\cite{liu2015deep} which is widely employed in the recent DeepFake studies~\cite{juefei2021countering}. The facial images in CelebA are employed for creating fake faces (\eg{}, attribute editing, face reenactment, identity swap) via various GANs. CelebA contains more than 200K facial images with $40$ attributes annotation for each face. All the facial images are cropped to 256$\times$256.

\noindent\textbf{Model Architectures.} Our proposed method is evaluated on three types of DeepFake that involve source images manipulation. For the attribute editing, StarGAN~\cite{choi2018stargan}, AttGAN~\cite{he2019attgan}, and Fader Network~\cite{lample2017fader} are employed for fine-grained facial attribute editing. For face reenactment, we employ the public available tool Icface~\cite{tripathy2020icface} to swap facial expressions. For identity swap, we adopt the popular DeepFake tool faceswap \cite{Faceswap} to swap faces freely.

\noindent\textbf{Baselines.} In evaluation, we compare our work with prior studies by employing gradient-based PGD and optimization-based strategy C\&W to generate imperceptible perturbations for a comprehensive comparison.

\noindent\textbf{Evaluation Metrics.} To evaluate the visual quality of manipulated facial images by injecting our proposed perceptual-aware perturbations, we adopt three different metrics, the average MSE, PSNR, and SSIM, for measuring the similarity between the original image and the disrupted fake image. Furthermore, we employ attack success rate (ASR) to report the successfully disrupted facial images, where the distortion measured by $L_2 \ge 0.05$.



\noindent\textbf{Implementation.} In our comparison experiments, the iteation for the PGD is $10$, the optimizer for C\&W and our method is Adam, the learning rate is $10^{-4}$, the iteration is $500$, the $\epsilon$ set to $0.05$. All the experiments were performed on a server running Red Hat 4.8 system on an 80-core 2.50 GHz Xeon CPU with 187 GB RAM and an NVIDIA Tesla V100 GPU with 32 GB memory.

\subsection{Effectiveness Evaluation}
We report the experimental results of effectiveness evaluation on all three types of DeepFakes involving the manipulation of source images. Since a large portion of GANs in image synthesis are fine-grained attribute editing, we employ three different GANs (\eg{}StarGAN, AttGAN, and Fader Network) which could represent the SOTA performance of GANs in attribute editing. Experimental results show that our proposed method by operating the Lab color space could disrupt the SOTA GANs for DeepFake creation with competitive performance. The  visualization of our proposed method in disrupting DeepFaked images refer to the supplemental material.

Table~\ref{Table:effectiveness} summarizes the effectiveness evaluation and comparison with two baselines. For the three attributes editing GANs, our method achieved competitive results with the two baselines measured by four different metrics and significantly outperforms them in AttGAN. In the metrics of ASR by giving a distortion restriction, our method outperforms the two baselines in the three GANs for attributes editing. 

To the best of our knowledge, this is the first work to present a comprehensive evaluation on the three types of DeepFakes, including attribute editing, identity swap, and face reenactment. Experimental results in Table~\ref{Table:effectiveness} demonstrated the effectiveness of our proposed method and illustrated that our proposed method achieved competitive results in comparing with the two baselines.

\begin{table*}[t]
\scriptsize
\centering
\setlength{\tabcolsep}{3.5pt}
\begin{tabular}{c|c|c|c|c|c|c|c|c|c|c|c|c}
\toprule
\multirow{2}{*}{\textbf{DeepFake Model}} & \multicolumn{4}{c|}{\textbf{PGD}} & \multicolumn{4}{c|}{\textbf{C\&W}} & \multicolumn{4}{c}{\textbf{Our method}}\\
&\textbf{$L_2\uparrow$}& \textbf{PSNR $\downarrow$}&\textbf{SSIM$\downarrow$}& \textbf{ASR$\uparrow$}&\textbf{$L_2\uparrow$}& \textbf{PSNR$\downarrow$}&\textbf{SSIM$\downarrow$}& \textbf{ASR$\uparrow$}&\textbf{$L_2\uparrow$}& \textbf{PSNR$\downarrow$}&\textbf{SSIM$\downarrow$}& \textbf{ASR$\uparrow$}\\ \midrule
StarGAN~\cite{choi2018stargan} & 1.192 & 5.386 & 0.297 & \textbf{100.0} & \textbf{1.409} & \textbf{4.781} & 0.285 & \textbf{100.0} & 1.302 & 5.292 & \textbf{0.251} & \textbf{100.0}\\
AttGAN~\cite{he2019attgan} & 0.091 & 16.885 & 0.701 & 80.50 & 0.095 & 16.419 & 0.679 & 84.10 & \textbf{0.103} & \textbf{15.975} & \textbf{0.634} & \textbf{85.30}\\
Fader Network~\cite{lample2017fader} & 0.195 & 13.663 & 0.374 & 89.70 & \textbf{0.242} & \textbf{10.932} & 0.365 & 91.40 & 0.229 & 11.047 & \textbf{0.320} & \textbf{93.10}\\
Identity swap & 0.035 & 15.976 & 0.468 & 5.40 & \textbf{0.031} & 16.710 & \textbf{0.513} & \textbf{3.90} & 0.029 & \textbf{17.153} & 0.530 & 4.20\\
Face reenactment~\cite{tripathy2020icface} & - & 18.336 & 0.688 & - & \textbf{-} & 16.621 & \textbf{0.627} & \textbf{-} & - & \textbf{17.824} & 0.640 & -\\
\bottomrule
\end{tabular}
\caption{Performance of our method in disrupting the three DeepFake types with a comparison with two competitive baselines. The manipulated attributes for the four GANs are Black hair, Blond hair, Brown hair, male, and young. For the adopted four metrics, the large value of $L_2$ and ASR indicates the large distortion is introduced while the small value of PSNR and SSIM means the large corruption is introduced. For the face reenactment, the ASR is applicable since the frames are extracted from the video, thus we leave it blank.}
\label{Table:effectiveness}
\end{table*}

\subsection{Robustness Analysis}
In this section, we report the evaluation results of our method against common input transformations, including JPEG compression, Gaussian Blur, reconstruction, \etc{}. Specifically, in the real world scenarios, the input with added perturbations would be spread in the social media platform and suffers various real degradations (\eg{}compression, blur, \etc{}). A recent study MagDR revealed that the prior DeepFake disruption studies failed in tackling input reconstruction where the added imperceptibly could be easily removed or corrupted by simple reconstruction. Thus, we aim to answer one question that \emph{whether our proposed method is robust against the common input transformations, especially the input reconstruction studied in MagDR.}

\textbf{Performance on common input transformations.} Table~\ref{Table:robustness} summarizes the performance of our method in tackling various input transformations. Experimental results show that our method shows promising performance in the adversary settings and demonstrated its robustness in surviving various input transformations. Specifically, our method significantly outperforms the two competitive baselines in all four adversary settings. In Table~\ref{Table:robustness}, the second row \emph{Defense} means a clean setting without any input transformations; the third row \emph{JPEG compression} indicates that the input is compressed before leveraged for creating DeepFakes; the fourth row \emph{JPEG Compression ($\epsilon=0.1$)} denotes that the added perturbations is restricted to a certain range larger than the default $\epsilon=0.05$; the fifth row \emph{Gaussian Blur} represents that Gaussian blur is employed to input; the last row \emph{Blur (Data augmentation)} explores whether the resistant against blur can be improved via data augmentation. Experimental results show that our method outperforms the two baselines measured by four different metrics on four input transformations.

\begin{table*}[tp]
\scriptsize
\centering
\setlength{\tabcolsep}{3.5pt}
\begin{tabular}{c|c|c|c|c|c|c|c|c|c|c|c|c}
\toprule
\multirow{2}{*}{\textbf{Defense}} & \multicolumn{4}{c|}{\textbf{PGD}} & \multicolumn{4}{c|}{\textbf{C\&W}} & \multicolumn{4}{c}{\textbf{Our method}}\\
&\textbf{$L_2\uparrow$}& \textbf{PSNR $\downarrow$}&\textbf{SSIM$\downarrow$}& \textbf{ASR$\uparrow$}&\textbf{$L_2\uparrow$}& \textbf{PSNR$\downarrow$}&\textbf{SSIM$\downarrow$}& \textbf{ASR$\uparrow$}&\textbf{$L_2\uparrow$}& \textbf{PSNR$\downarrow$}&\textbf{SSIM$\downarrow$}& \textbf{ASR$\uparrow$}\\ \midrule
No defense & 1.192 & 5.386 & 0.297 & \textbf{100.0} & \textbf{1.409} & \textbf{4.781} & 0.285 & \textbf{100.0} & 1.302 & 5.292 & \textbf{0.251} & \textbf{100.0}\\
JPEG compression & 0.038 & 17.236 & 0.682 & 41.30 & 0.043 & 16.774 & 0.611 & 45.20 & \textbf{0.058} & \textbf{12.722} & \textbf{0.433} & \textbf{54.10}\\
JPEG Compression ($\epsilon=0.1$) & 0.568 & 9.430 & 0.484 & 83.50 & 0.501 & 11.355 & 0.552 & 81.20 & \textbf{0.722} & \textbf{7.459} & \textbf{0.370} & \textbf{90.70}\\
Gaussian Blur ($\delta=3$) & 0.007 & 28.086 & 0.944 & 0.0 & 0.004 & 29.335 & 0.913 & 0.0 & \textbf{0.012} & \textbf{20.252} & \textbf{0.865} & \textbf{5.80}\\
Blur (Data Augmentation) & 0.096 & 26.871 & 0.806 & 77.90 & 0.114 & 21.335 & 0.713 & 81.30 & \textbf{0.302} & \textbf{18.252} & \textbf{0.631} & \textbf{88.50}\\
\bottomrule
\end{tabular}
\vspace{-5pt}
\caption{Performance of our method on StarGAN in adversary settings with a comparison with two competitive baselines. The inputs are transformed by blurring, input compression.}
\label{Table:robustness}
\vspace{-5pt}
\end{table*}

\textbf{Performance on MagDR~\cite{chen2021magdr}}. Beyond the aforementioned popular input transformations, a recent study MagDR revealed that existing DeepFake disruption techniques are not robust to input reconstruction. Specifically, the added imperceptible perturbations could be corrupted or even removed by employing simple reconstruction and further output a high-quality DeepFake without noticeable artifacts. In our experiments, we also explore whether our method could resist the threat of input reconstruction.

Specifically, Table~\ref{Table:magdr} summarizes the experimental results and compares them with one baseline. We reproduce MagDR and ensure the implementation details are correct by checking with the authors of MagDR. For the input with added perturbations, our method exhibits fewer perturbations than the baseline but leaves more perturbations in the input after employing MagDR which is our desired result as it would be lead to corrupted DeepFakes. For the output by employing MagDR, the value of the three metrics indicates the effectiveness in creating DeepFakes with noticeable artifacts. We can find that the MagDR failed in corrupting our added perceptual-aware perturbations over the three metrics by operating the Lab color space in comparison with the baseline.
\vspace{-5pt}

\begin{table}[tp]
\scriptsize
\centering
\setlength{\tabcolsep}{3.5pt}
\begin{tabular}{c|c|c|c|c}
\toprule
Metric & PGD & Our Method & PGD (MagDR) & Our Method (MagDR)\\ \midrule
SSIM (I) & 0.914 & \textbf{0.953} & 0.914 & \textbf{0.906}\\
PSNR (I) & 35.313 & \textbf{35.618} & 35.356 & \textbf{30.699}\\
MSE (I) & 29.836 & \textbf{27.889}  & 29.546 & \textbf{88.320}\\
\bottomrule\bottomrule
SSIM (O) & 0.604 & \textbf{0.522} & 0.892 & \textbf{0.819}\\
PSNR (O) & 15.029 & \textbf{13.107} & 30.838 & \textbf{20.828}\\
MSE (O) & 4117.143 & \textbf{6795.289} & 85.429 & \textbf{1083.934} \\
\bottomrule
\end{tabular}
\caption{Performance of our method in bypassing input reconstruction. SSIM (I) indicates the input with added perturbations, while SSIM (O) means the output of SSIM (I) by employing DeepFake manipulations. PGD (MagDR) and our method (MagDR) represent the sample is reconstructed by prior study MagDR.}
\label{Table:magdr}
\end{table}

In summary, a comprehensive robustness evaluation against both the common input transformations and input reconstruction demonstrated that our proposed method shows strong capabilities in resisting common input transformations and outperforms the two baselines measured by four different metrics in tackling the three common input transformations. Our proposed method shows its potential to be deployed in the real-world scenarios.

\subsection{Performance across Diverse GANs}
In this section, we explore whether our generated perturbations on one model are effective on other GAN models as well. Table~\ref{Table:generic} summarizes the experimental results of our proposed method in tackling diverse GANs in black-box settings. Specifically, our method outperforms  PGD except one case where the perturbations are generated from Fader Network and apply to StarGAN. However, the other baseline C\&W achieved the best performance in almost all the cases. It should be noted that the ASR value of our method is similar to C\&W, thus our method also has a good transferability across GANs. Thus, it would be interesting to explore perturbations with strong transferability across diverse GANs, especially to combine C\&W for achieving both high transferability and robustness against input transformation attack, which is our future work.

\begin{table}[t]
\scriptsize
\centering
\setlength{\tabcolsep}{2.5pt}
\begin{tabular}{c|c|c|c|c|c|c|c|c|c}
\toprule
\multirow{2}{*}{\textbf{GAN}} & \multicolumn{3}{c|}{\textbf{StarGAN}} & \multicolumn{3}{c|}{\textbf{AttGAN}} & \multicolumn{3}{c}{\textbf{Fader Network}}\\
&\textbf{PGD}& \textbf{C\&W}&\textbf{Our}& \textbf{PGD}&\textbf{C\&W}& \textbf{Our}&\textbf{PGD}& \textbf{C\&W}&\textbf{Our}\\ \midrule
StarGAN & - & - & - & 7.1 & 11.5 & \textbf{13.6} & 9.7 & \textbf{16.8} & 15.3\\
AttGAN & 26.3 & \textbf{37.1} & 35.4 & - & - & - & 18.4 & \textbf{21.5} & 19.6\\
Fader Network & \textbf{16.2} & 20.7 & 18.9 & 5.3 & \textbf{7.8} & 7.0 & - & - & -\\
\bottomrule
\end{tabular}
\caption{Performance of our method in tackling a diverse GANs. The performance is evaluated by employing ASR. Our indicates our proposed method.}
\label{Table:generic}
\end{table}

\subsection{Exploring other Color Spaces}
To better illustrate the advances by operating on the Lab color space, we investigate the other three popular color space, namely RGB, HSV, and CMYK to explore their performance in disrupting DeepFakes and their stealthiness in evading detection. Experimental results in Table~\ref{Table:color_space} show that our proposed method operating on the Lab color space outperforms the other three baselines measure by $L_2$ in resisting the two types of input transformation (\eg{}compression and Gaussian blur), which exposes more visually artifacts.

\begin{table}[t]
\scriptsize
\centering
\setlength{\tabcolsep}{3.5pt}
\begin{tabular}{c|c|c|c|c}
\toprule
Defense&\textbf{RGB}&\textbf{Lab}&\textbf{HSV}&\textbf{CMYK}\\ \midrule
JPEG Compression & 0.038 & \textbf{0.058} & 0.053 & 0.041  \\
Gaussian Blur ($\sigma$=1)  & 0.031 & \textbf{0.049} & 0.040 & 0.038 \\
Gaussian Blur ($\sigma$=2)  & 0.016 & \textbf{0.025} & 0.019 & 0.013 \\
Gaussian Blur ($\sigma$=3)  & 0.007 & \textbf{0.012} & 0.008 & 0.007 \\
\bottomrule
\end{tabular}
\caption{Performance of our method operating on the Lab color space on StarGAN in an adversary settings with a comparison with other three color space. The input are tranformed by input compression and blurring.}
\label{Table:color_space}
\end{table}

To further explore the stealthiness of the perturbations generated by operating the Lab color space, we employ an adversarial noise detector by using local intrinsic dimensionality for detection~\cite{ma2018characterizing}. Experimental results in Table~\ref{Table:auc_color_space} shows that our proposed method is more stealthy than the other three baselines with lower AUC score.

\begin{table}[t]
\scriptsize
\centering
\setlength{\tabcolsep}{3.5pt}
\begin{tabular}{c|c|c|c|c}
\toprule
GAN&\textbf{RGB}&\textbf{Lab}&\textbf{HSV}&\textbf{CMYK}\\ \midrule
StarGAN & 89.51 & \textbf{85.43} & 88.52 & 88.73  \\
AttGAN  & 91.53 & 84.60 & \textbf{82.53} & 89.37 \\
Fader Network  & 92.31 & 87.33 & 87.15 & \textbf{86.46} \\
\bottomrule
\end{tabular}
\caption{Performance of our method operating on the Lab color space in evading noise detection comparison with three baselines from three different color space.}
\label{Table:auc_color_space}
\end{table}

\subsection{Ablation Study}
\begin{figure}[h]
\centering
\includegraphics[width=0.75\linewidth]{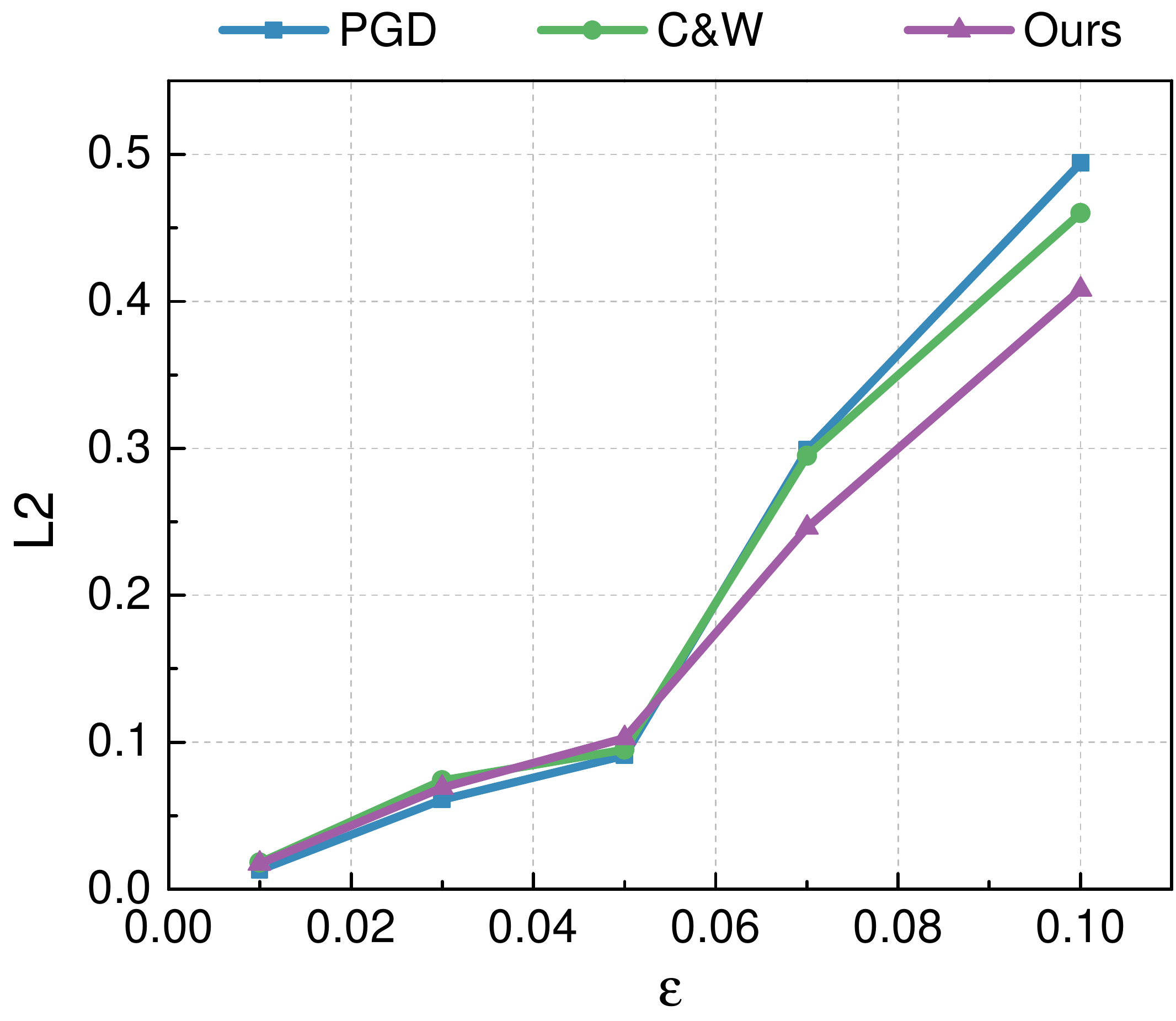}
\caption{The trend of the magnitude of perturbations $\epsilon$ and the intensity of damaged DeepFake.}
\label{fig:magnitude}
\end{figure}

In our ablation study, we explore the trend our added perturbations $\epsilon$ and the degree of damaged DeepFakes in comparison with two baselines. Figure~\ref{fig:magnitude} shows us the manipulated images with StarGAN by adding the perturbation $\epsilon$ from $0.01$ to $0.1$ with a default interaction $500$. Experimental results shown that a large added perturbation leads to large distortion in the created DeepFakes, however large perturbations also introduce unnatural artifacts into the source data which will break our ``natural" property requirement. Thus, we set the maximum perturbation to $0.05$ in our whole experiments to ensure a fair comparison with the two baselines.

On the one hand, our generated perturbations for disrupting DeepFakes by exposing noticeable artifacts to avoid users believe the misinformation caused by such DeepFakes, on the other side, we hope that the disrupted DeepFakes could provide strong fake signals for the simple classifier. Table~\ref{Table:cnn} summaries the performance of a simple and popular DeepFake detector~\cite{wang2020cnn} in spotting unknown DeepFakes. Wang \etal{}'s~\cite{wang2020cnn} study achieved merely 70.69\% in spotting images manipulated by StarGAN while the source image is clean.  Wang \etal{} give an accuracy more than 87\% in spotting our disrupted images by adding our perceptual-aware perturbations into the source image where the magnitude of the perturbation is only $0.01$. Thus, our proposed method show promising results in providing clear fake textual signals for DeepFake detectors in tacking unknown DeepFakes.

\begin{table}[h]
\scriptsize
\centering
\setlength{\tabcolsep}{3.5pt}
\begin{tabular}{c|c|c|c|c}
\toprule
Clean & \makecell[c]{iter =10 \\ ($\epsilon=0.01$)} & \makecell[c]{iter =100 \\ ($\epsilon=0.01$)} & \makecell[c]{iter =100 \\ ($\epsilon=0.03$)} & \makecell[c]{iter =300 \\ ($\epsilon=0.03$)}\\ \midrule
70.69\% & 75.52\% & \textbf{87.81\%} & 85.54\% & 80.70\% \\
\bottomrule
\end{tabular}
\caption{The performance of Wang \etal{}in spotting unknown DeepFakes.}
\label{Table:cnn}
\end{table}

\begin{figure}[t]
\centering
\includegraphics[width=1\linewidth]{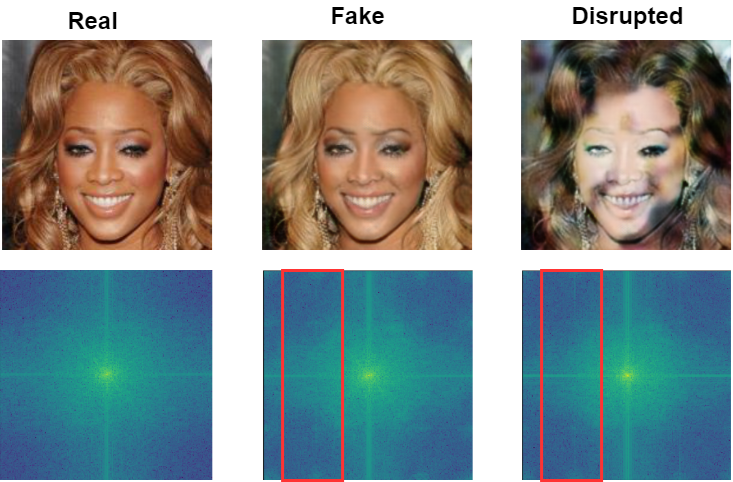}
\caption{In the first row, the images in turn are a real image from CelebA, a fake image produced by StarGAN on the clean real image, a disrupted image produced by StarGAN on a image added our perceptual-aware perturbations. In the second row, the images are the spectrum corresponding to the images above.}
\label{fig:signals}
\end{figure}

Additionally, we also present Figure~\ref{fig:signals} to visualize whether our proposed method could enhance the fake textual. In Figure~\ref{fig:signals}, the red rectangle highlights that our created DeepFakes from the images with our added perceptual-aware perturbations exhibit obvious fake textual in spectrum than the fake image manipulated on clean image.

\subsection{Visualization}
\begin{figure}[t]
\centering
\includegraphics[width=1\linewidth]{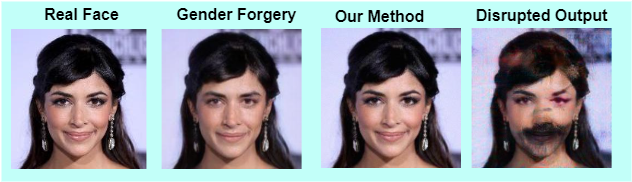}
\caption{Visualization of our proposed anti-fogery method. The first column is the real face and the second column indicates the forgery face by manipulating gender. The third column indicates the face by adding our perceptual-aware perturbations with high visually quality and the last column represents the disrupted output by manipulating the faces with our added perturbations.}
\label{fig:visual}
\end{figure}

Figure~\ref{fig:visual} presents the visualization of our proposed method in disrupting DeepFaked images.

\subsection{Discussion}
Here, we discuss the limitations of our proposed method by operating on the Lab color space to generate perceptual-aware perturbations for DeepFake disruption and suggest provisional countermeasures against them. Our generated perturbations are added into the whole input which suffers the potential threat that the adversary crops the background to destroy the perturbations directly. In our future work, we explore more robust technique to evade such attack by setting masks to enforce the perturbations are added into the internal region of face.

\section{Conclusion and Future Research Direction}\label{sec:con}
We propose the first adversarial perceptual-aware perturbations operating on the Lab color space for anti-fogery and performed an extensive evaluation of our method on three types of DeepFakes. Experimental results demonstrate its effectiveness in disrupting DeepFakes by exposing noticeable artifacts to human eyes while preserving the visually naturalness of the source image. More importantly, our method exhibits its robustness against input reconstruction which significantly outperforms prior studies. In addition, the created DeepFakes by adding our perceptual-aware perturbations provide strong signals for DeepFake detection. Our findings of perceptual-aware adversarial perturbations present a new insight for defending DeepFakes proactively by investigating more natural and robust adversarial perturbations like facial background~\cite{xiao2020noise}, \etc{}. Thus, in our future work, we will explore more physical scenes to serve as guards for protecting images being DeepFaked.

\section{Broader Impact} \label{sec:impact}
With the rapid development of GAN in image synthesis and the popularity of social media in sharing the exciting moments with personal photos, DeepFake is becoming a real threat to individuals and celebrities as the potential of creating fake pornography and releasing fake statements. The community seeks various countermeasures to fight DeepFakes in both passive and proactive defense manner, unfortunately both of them are still in its fancy and not prepared for tackling this emerging threat. The passive DeepFake detection is an ex-post forensics method while the existing studies for proactive DeepFake disruption are not robust to input reconstruction.

To address the challenges in DeepFake defense, our work is the first attempt to identify and showcase that adversarial perturbations by operating on the Lab color space is not only feasible, but also leads to a robust protection of facial images without introducing visually artifacts. In a large sense, this work can and will provide new thinking into how to better design robust anti-forgery techniques for defending DeepFakes in the wild in order to mitigate the security and privacy concerns caused by the spreading of DeepFakes.

\section*{Acknowledgments}
This research was supported in part by the National Key Research and Development Program of China under No.2021YFB3100700, the Fellowship of China National Postdoctoral Program for Innovative Talents under No.BX2021229, the Fundamental Research Funds for the Central Universities under No.2042021kf1030, the Natural Science Foundation of Hubei Province under No.2021CFB089, the Open Foundation of Henan Key Laboratory of Cyberspace Situation Awareness under No. HNTS2022004, the National Natural Science Foundation of China (NSFC) under No.61876134, No.U1836202, No.62076187, No.62172303.



\footnotesize
\bibliographystyle{named}
\bibliography{ref}


\end{document}